\documentstyle[11pt,aaspp4]{article}

\lefthead{Suginohara, Suginohara, \& Spergel}
\righthead{Cross-Correlating CMB with Redshift Surveys}

\begin{document}

\title{Cross-Correlating Cosmic Microwave Background\\
Radiation Fluctuations with Redshift Surveys:\\
Detecting the Signature of Gravitational Lensing}

\author{Maki Suginohara\altaffilmark{1,2},
Tatsushi Suginohara\altaffilmark{1,3,4},
and David N. Spergel\altaffilmark{5,6}}
\affil{Department of Astrophysical Sciences, Princeton University,
Princeton, NJ 08544}

\altaffiltext{1}{Department of Physics, the University of Tokyo,
Tokyo 113, Japan}
\altaffiltext{2}{makis@astro.Princeton.EDU, JSPS Research Fellow}
\altaffiltext{3}{RESCEU, School of Sciences, the University of Tokyo,
Tokyo 113, Japan}
\altaffiltext{4}{tatsushi@astro.Princeton.EDU, JSPS Postdoctoral Fellow}
\altaffiltext{5}{Department
of Astronomy, University of Maryland, College Park, MD 20742}
\altaffiltext{6}{dns@astro.Princeton.EDU}

%
%

\begin{abstract}
Density inhomogeneities along the line-of-sight distort fluctuations
in the cosmic microwave background. Usually, this effect is thought of
as a small second-order effect that mildly alters the statistics of
the microwave background fluctuations. We show that there is a
first-order effect that is potentially observable if we combine
microwave background maps with large redshift surveys. We introduce a
new quantity that measures this lensing effect, $<T(\delta
\mbox{\boldmath $\theta \cdot \nabla $}T)>,$ where $T$ is the
microwave background temperature and $\delta \mbox{\boldmath $\theta
$}$ is the lensing due to matter in the region probed by the redshift
survey. We show that the expected signal is first order in the
gravitational lensing bending angle, $<(\delta \mbox{\boldmath $\theta
$})^2>^{1/2},$ and find that it should be easily detectable,
(S/N)$\sim 15-35,$ if we combine the Microwave Anisotropy Probe
satellite and Sloan Digital Sky Survey data. Measurements of this
cross-correlation will directly probe the ``bias'' factor, the
relationship between fluctuations in mass and fluctuations in galaxy
counts.
\end{abstract}



\clearpage
%
%
\section{Introduction}

The next several years should be very exciting for cosmologists:
Microwave Anisotropy Probe (MAP; \cite{wri96}) and PLANCK
(\cite{bou95}) will make high resolution maps of the microwave
background sky; while the Sloan Digital Sky Survey (SDSS;
\cite{gun95}; see also http://www-sdss.fnal.gov:8000/) will measure
redshifts of 10$^6$ galaxies and positions of 10$^8$ galaxies. In this
paper, we explore the {\it direct\/} connection between these two
measurements through gravitational lensing: the path of a cosmic
microwave background (CMB) photon is distorted by inhomogeneities in
the matter distribution; galaxy surveys detect these inhomogeneities
as fluctuations in galaxy number counts.

The effect of the gravitational lensing on the CMB anisotropies has
been studied by many authors. The uncomfortably low upper limits
(\cite{uso84,rea89}) provoked a great deal of controversy
(\cite{kas88,tom88,sas89,wat91}) about the possibility that
gravitational lensing washes out the intrinsic fluctuation. After the
detection by the Cosmic Background Explorer (COBE; \cite{smo92}),
there has been renewal of interest (\cite{lin90a}, b, \cite{cay93a},
b, \cite{sel96}) in investigating how the CMB power spectrum is
redistributed owing to gravitational lensing. For example, Seljak
(1996) recently presented detailed calculations of gravitationally
deflected CMB power spectra, including the effect of the nonlinear
evolution of matter density fluctuations. His result shows, however,
that the modification of the CMB power spectrum is a second-order
effect of the photon bending angle and less than a few percent on
angular scales greater than ten arcminutes. Hence, the lensing effect
on the CMB spectrum itself is extremely difficult to detect, even with
observations such as the MAP project.
Linder (1997) has also studied the effects of lensing on the
correlation function and has introduced a cross-correlation function
similar to the one that we study here.

In this paper, we introduce a cross-correlation function that is
sensitive to the gravitational lensing correlations between the
temperature fluctuations and matter density fluctuations. We show that
the cross-correlation is first-order in the bending angle so it should
be easier to detect if we have both accurate CMB maps and redshift
surveys. We quantitatively estimate its magnitude and its cosmic
variance in cold dark matter (CDM) universes. The rest of the paper is
organized as follows. We review the formalism developed by Seljak
(1996) for computing the angular excursion of the CMB photon paths on
celestial sphere in section~2. In section~3, we formulate the
cross-correlation between matter density inhomogeneities and CMB
temperature fluctuations. Section~4 concludes.

%
%

\section{Gravitational Lensing}

In this section, we review gravitational lensing by density
fluctuations. We follow the power spectrum approach of Seljak (1994,
1996). We focus on the angular excursions produced by matter
fluctuations at low redshifts, where they can be most easily inferred
from redshift surveys.

Fluctuations in matter density, $\delta $, generate variations in the
gravitational potential,
\begin{equation}
  \nabla ^2\phi =4\pi G\rho _ba^2\delta ,
\end{equation}
where $G$ is the gravitational constant, and $\rho _b$ is the mean
background mass density. Conventionally, the matter density
fluctuations are related to the fluctuations in galaxy counts by a
linear biasing parameter, $b$:
\begin{equation}
  \delta _g=b\,\delta .
\end{equation}
Since most of the lensing effects will be produced by fluctuations on
large physical scales ($k<0.1h^{-1}$Mpc), the linear biasing model
will hopefully be valid.
It is important to note that detailed nonlinear and/or time-dependent
biasing may somewhat change the statistics we present in this paper.

A photon emitted at some angular position $\mbox{\boldmath $\theta $}$
has been deflected by gravitational lensing during its long travel,
with the result that it is observed at different angular position,
$\mbox{\boldmath $\psi $}$. The photon angular excursion on celestial
sphere is given by Seljak (1994):
\begin{equation}
  \mbox{\boldmath $\theta $}-\mbox{\boldmath $\psi $} =\delta
  \mbox{\boldmath $\theta $}(z) =-2\int_0^{\chi (z)}d\chi ^{\prime }
  \,W(\chi ^{\prime },\chi _{\rm dec} )\mbox{\boldmath$\nabla
  $}_{\perp } \phi ,
\end{equation}
where $\mbox{\boldmath$\nabla $}_{\perp }$ is transverse component of
the potential gradient with respect to the photon path,
\begin{equation}
  W(\chi ,\chi _{\rm dec})=\sin _K(\chi _{\rm dec}-\chi )/
  \sin _K(\chi_ {\rm dec})
\label{eqn:pro}
\end{equation}
is a projection operator on celestial sphere, and $\chi _{\rm
dec}=\chi (z_{\rm dec})$ is unperturbed comoving radial distance
corresponding to redshift $z_{\rm dec}$ at decoupling time.  In
equation (\ref{eqn:pro}), $\sin _K(u)=\sin (u),u,$ and sinh(u) in a
closed, flat, and open universe, respectively.

Next, we consider the relative angular excursion $\delta
\mbox{\boldmath $\theta $}-\delta \mbox{\boldmath $\theta $}^{\prime }
$ of a photon pair emitted from angular positions $\mbox{\boldmath
$\theta $}$ and $ \mbox{\boldmath $\theta $}^{\prime }$.  We restrict
our calculation to the small angular separation limit, $\xi
=|\mbox{\boldmath $\theta $}-\mbox{\boldmath $\theta $} ^{\prime }|\ll
1$, and assume that the relative angular excursion $\delta
\mbox{\boldmath $\theta $}-\delta \mbox{\boldmath $\theta $}^{\prime }
$ obeys Gaussian statistics.  Lensing is primarily due to scattering
events from mass fluctuations on the 10 - 100 Mpc scale.  As there are
30 - 300 of these fluctuations between the surface of last scatter and
the present along each photon path, the central limit theorem implies
that this is a good approximation.  Following Seljak (1994), we
characterize the statistics of the lensing fluctuations by its
root-mean-square dispersion:
\begin{eqnarray}
  \sigma (\xi ;z) &=&2^{-1/2}\left\langle \left[ \delta
  \mbox{\boldmath $\theta $}(z)-\delta \mbox{\boldmath $\theta
  $}^{\prime }(z) \right]^2\right\rangle _\xi ^{1/2} =\left[ C_{{\rm
  gl}}(0;z)-C_{{\rm gl}}(\xi;z)\right] ^{1/2},
  \label{eqn:cgf} \\
  C_{{\rm gl}}(\xi ;z) &\equiv &\frac{2}{\pi}\int_0^\infty
  k^3dk\,\int_0^{\chi (z)}d\chi ^{\prime}\, P_\phi (k,\tau _0-\chi
  ^{\prime })W^2(\chi^{\prime },\chi _{\rm dec}) J_0(k\xi \sin _K\chi
  ^{\prime }), \nonumber
\end{eqnarray}
where $\langle \;\rangle _\xi $ denotes the averaging over pairs
observed with fixed angular separation $\xi$, $J_0$ is the Bessel
function of order $0$, and $P_\phi (k)$ is the gravitational potential
power spectrum. The power spectrum of the potential fluctuations are
related to the power spectrum of the density fluctuations through,
\begin{equation}
  P_\phi (k,\tau )=(9/4)\Omega _m^2(\tau )H^4(\tau )a^4(\tau )
  k^{-4}P(k,\tau ),
\end{equation}
where $\Omega _m$ is the mass density parameter given by $\Omega _
m\equiv 8\pi G\rho _b/(3H^2)$, and $H(\tau )$ is the Hubble parameter.

Figures 1a shows $\sigma (\theta ;z)$ as functions of $\theta $ for
several redshift values. We consider throughout this paper two
cosmological models: one is the standard CDM (SCDM) model with $\Omega
_{m0}=1$, $\Omega _{v0}=0$, $h=0.5$, and $\sigma _8=1.2$, and the
other is a low-density, cosmological constant dominated CDM ($\Lambda
$CDM) model with $\Omega _{m0}=0.3$, $\Omega _{v0}=0.7$, $h=0.7$, and
$\sigma _8=1.0,$ the best fit model of Ostriker and Steinhardt (1995). 
Here, $\Omega _{m0}$ and $\Omega _{v0}$ are the present mass density
and the present vacuum energy density normalized by the critical
density; $h$ is the present Hubble parameter in units of $100\, {\rm
km\,s^{-1}Mpc^{-1}}$; $\sigma _8$ is the mass fluctuations within a
sphere of radius $8\,h^{-1}\,{\rm Mpc}$. We use the COBE normalized
value (\cite{bun97}) for $\sigma _8$. In numerical calculation of
$\sigma (\xi ;z)$, we have used the fitting formula for CDM linear
transfer function given in Bardeen et al. (1986).

%
\begin{figure}[t]
  \plottwo{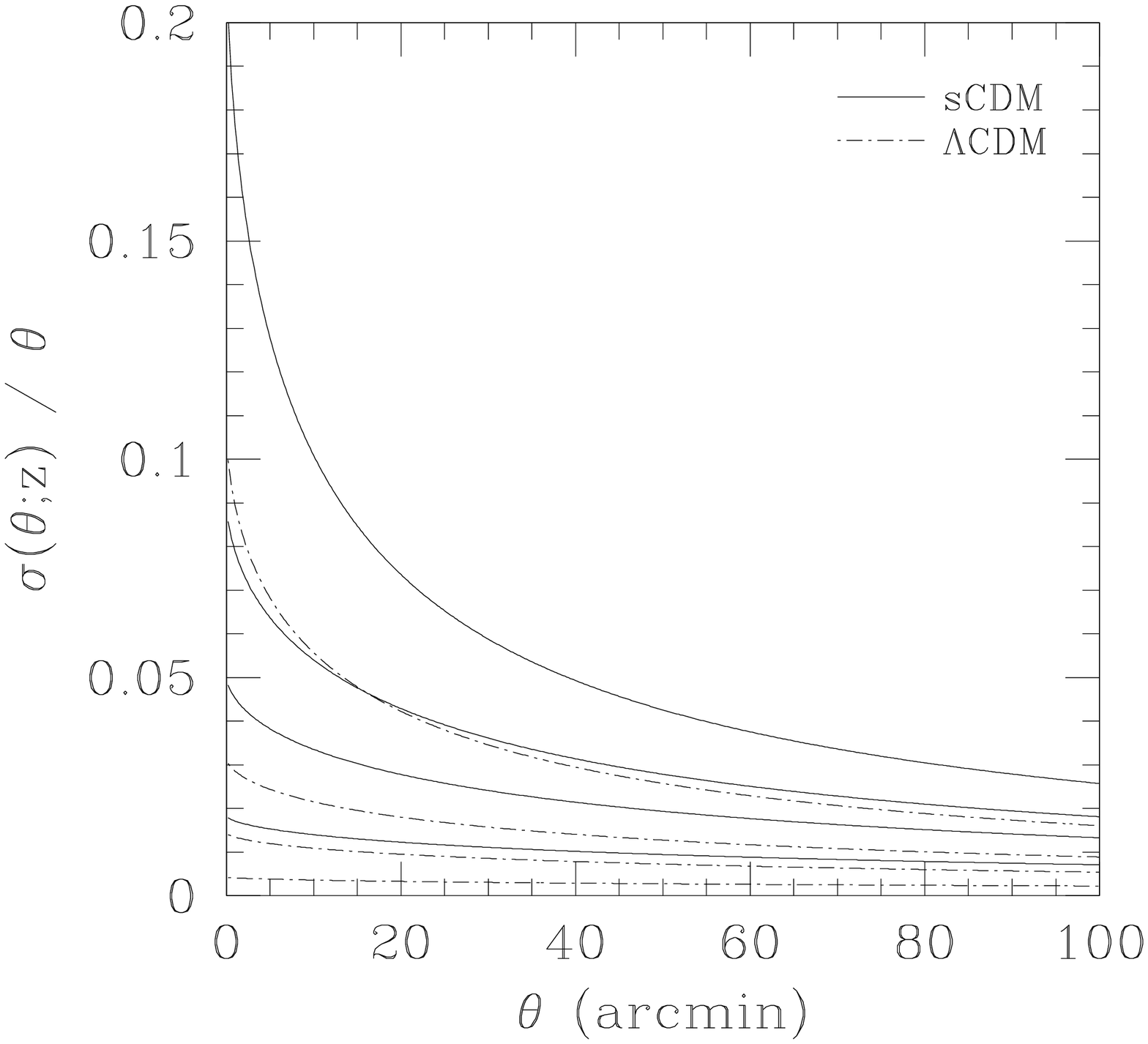}{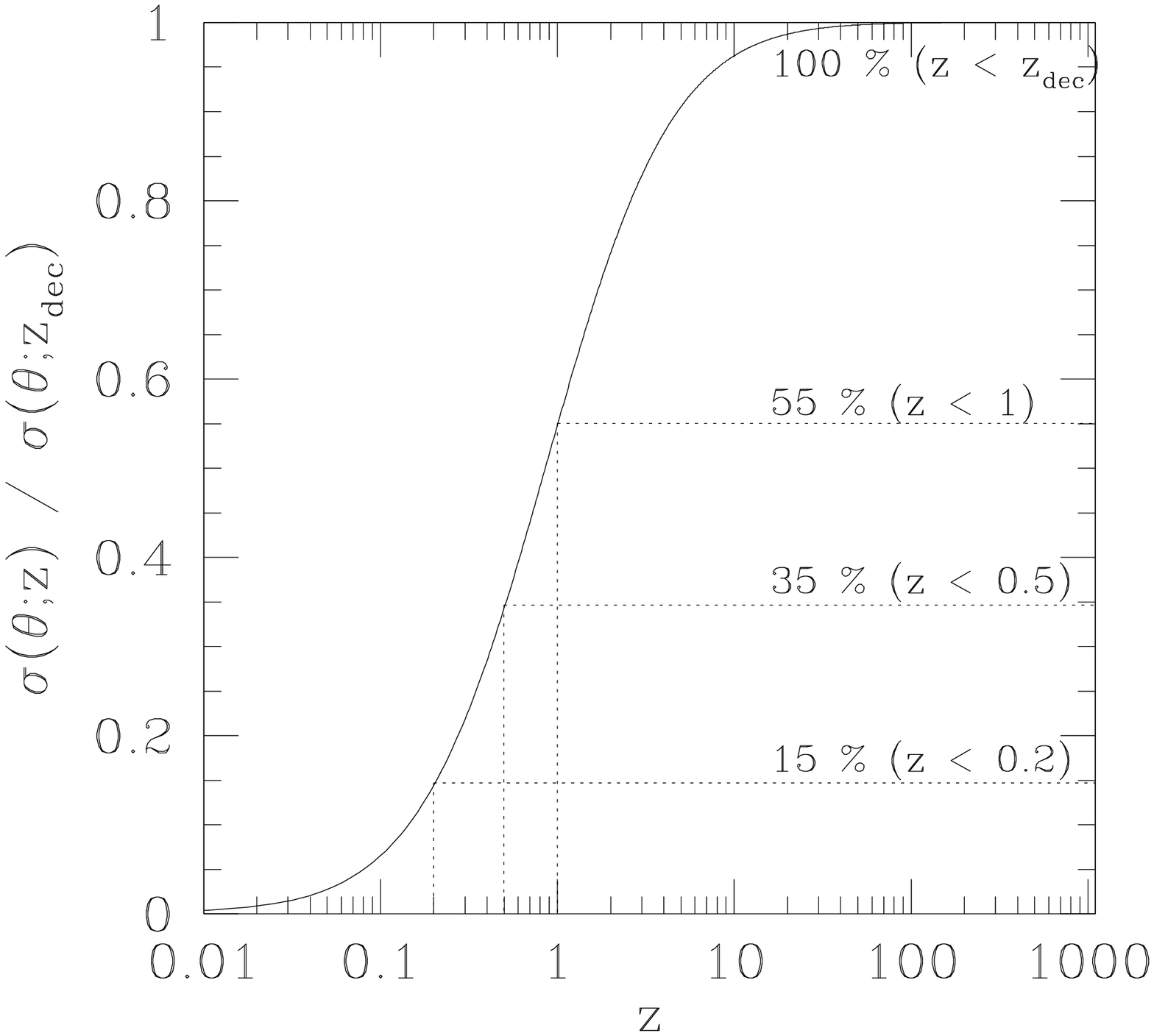}
  \caption{The dispersion of the relative
     angular excursion, $\sigma(\theta;z)$.  1a: Dependence on the
     separation angle $\theta $. Solid lines show the standard cold
     dark matter (SCDM) model with $\Omega_{m0}=1$, $\Omega_{v0}=0$,
     $h=0.5$, and $\sigma_8=1.2$.  Broken lines show the cosmological
     constant dominated cold dark matter ($\Lambda$CDM) model with
     $\Omega_{m0}=0.3$, $\Omega_{v0}=0.7$, $h=0.7$, and
     $\sigma_8=1.0$.  For each model, the curves show the case in
     which $z = z_{\rm dec}$, $1$, $0.5$, and $0.2$, from upper to
     lower.  1b: Dependence on the redshift $z$ at $\theta=0.21\, {\rm
     degree}$ in the SCDM model.}
  \label{fig01ab}
\end{figure}

The SDSS project measures the redshift of galaxies up to $z\simeq 0.2$
within solid angle $\Omega _{\rm SDSS}=\pi \,{\rm steradian}$.
Furthermore, we expect to obtain photometric redshifts (\cite{con95})
of galaxies up to $z\simeq 1$ with fairly small uncertainties. Then we
will obtain the galaxy number density perturbation, $\delta _
g(\mbox{\boldmath $r$})$, within solid angle $\Omega _{\rm SDSS}$ for
$z<0.2$, and for $z<1$ with some uncertainties due to error in
photometric redshifts. We can see from Figures 1b that the
matter within $z<0.2$, $z<0.5$, and $z<1$ contributes 15\%, 35\%, and
55\%, respectively, to the angular excursion from the last scattering
surface. Hereafter, the values with subscript $A$ denote the integral
from the observer to some redshift $z$, the edge of the survey data,
and those with subscript $B$ denote the integral over remaining part;
for example:
\begin{eqnarray}
  \delta \mbox{\boldmath $\theta $}_A &=&-2\int_0^{\chi (z)}d\chi
  ^{\prime }\,W(\chi ^{\prime },\chi _{\rm dec}) \mbox{\boldmath
  $\nabla $}_{\perp }\phi , \\
  \delta \mbox{\boldmath $\theta $}_{B} &=&-2\int_{\chi (z)}^{\chi _
  {\rm dec}}d\chi ^{\prime }\, W(\chi ^{\prime },\chi _{\rm
  dec})\mbox{\boldmath $\nabla $}_{\perp }\phi .
\end{eqnarray}
The total lensing deviation, $\delta \mbox{\boldmath $\theta $}=
\delta \mbox{\boldmath $\theta $}_A+\delta \mbox{\boldmath $\theta
$}_B.$ Its variance is the sum of the contribution from the two
regions:
\begin{equation}
  \left\langle \left( \delta \mbox{\boldmath $\theta $}\right)
  ^2\right\rangle =\left\langle \left( \delta \mbox{\boldmath $\theta
  $}_A\right) ^2\right\rangle +\left\langle \left( \delta
  \mbox{\boldmath $\theta $} _B\right) ^2\right\rangle \equiv 2(\sigma
  _ A^2+\sigma _B^2).
\end{equation}
For our purposes, the distant lensing is unimportant; its only effect
is to slightly reduce the amplitude of the temperature fluctuations.

%
%

\section{Lensing the Microwave Background}

Gravitational lensing distorts the microwave background sky:
\begin{equation}
  \widetilde{T}(\mbox{\boldmath $\psi $})=T\left[ \mbox{\boldmath
  $\theta $}( \mbox{\boldmath $\psi $})\right] =T\left(
  \mbox{\boldmath $\psi $}+ \delta \mbox{\boldmath $\theta $}\right)
  =T(\mbox{\boldmath $\psi $})+ \delta \mbox{\boldmath $\theta \cdot
  \nabla $}T( \mbox{\boldmath $\psi $})+ \frac{1}{2}\left[ \delta
  \mbox{\boldmath $\theta \cdot \nabla $} \right]^2 T\left(
  \mbox{\boldmath $\psi $} \right) + \cdots .
\end{equation}
Here $\widetilde{T}$ denotes the measured temperature map and $T$
denotes the unlensed temperature map.

This distortion alters the statistics of the microwave background by
smearing out temperature correlations (Seljak 1996):
\begin{eqnarray}
  \left\langle \widetilde{T}(\mbox{\boldmath $\psi $})
  \widetilde{T}(\mbox{\boldmath $\psi $}^{\prime })\right\rangle & =
  &\left\langle T(\mbox{\boldmath $\psi $}) T(\mbox{\boldmath $\psi
  $}^{\prime })\right\rangle + \left\langle \left[ \delta
  \mbox{\boldmath $\theta \cdot \nabla$} T(\mbox{\boldmath $\psi
  $})\right] \left[ \delta \mbox{\boldmath $\theta $}^{\prime}
  \mbox{\boldmath $\cdot \nabla$} T(\mbox{\boldmath $\psi $}^{\prime })
  \right] \right\rangle \nonumber \\
  & & \qquad +\left\langle T(\mbox{\boldmath $\psi $}) (\delta
  \mbox{\boldmath $\theta $}^{\prime } \mbox{\boldmath $\cdot
  \nabla$})^2T (\mbox{\boldmath $\psi $}^{\prime })\right\rangle .
\end{eqnarray}
This can be alternatively written in terms of the correlation
function:
\begin{equation}
  C_{\rm lensed}(\xi )=C_{\rm unlensed}(\xi )+\frac{\sigma ^2}{2} \,
  \frac{\partial ^2C_{\rm unlensed}(\xi )}{\partial \xi ^2},
\end{equation}
where $\xi =\left| \mbox{\boldmath $\psi $}-\mbox{\boldmath $\psi
$}^{\prime } \right| .$

If we have a redshift survey, then the effects of gravitational
lensing of the microwave background can be observed more easily. Here,
we introduce a new quantity, $H,$ which measures the cross-correlation
between the temperature map and the predicted lensing:
\begin{equation}
  H\left( \xi \right) ={\cal N} \left\langle
  \widetilde{T}(\mbox{\boldmath $\psi $} )\left( \left[
  \delta\mbox{\boldmath$\theta $}_A- \delta \mbox{\boldmath $\theta
  $}_A^{\prime }\right] \mbox{\boldmath $\cdot \nabla
  $}\widetilde{T}(\mbox{\boldmath $\psi $} ^{\prime })\right)
  \right\rangle _\xi ,
\end{equation}
where $\delta \mbox{\boldmath $\theta $}_A(\mbox{\boldmath $\psi $})$
is determined from the redshift survey, and ${\cal N}$ is a
normalization factor given by
\begin{equation}
  {\cal N}^{-1} \equiv
  \sigma_A(\theta _{\rm fwhm})
  \left \langle T^2 \right \rangle _ {\theta _{\rm fwhm}}^{1/2} \left
  \langle \left( \mbox{\boldmath $\nabla $}T \right)^2 \right \rangle _
  {\theta _{\rm fwhm}}^{1/2}
\end{equation}
at a typical value of angular separation $\theta = \theta_{\rm fwhm}$.

We can evaluate this by expanding the temperature map in a Fourier
series.  Since the lensing bending angles are small, we simplify the
equations by making the plane parallel approximation and expanding out
the unlensed temperature map:
\begin{equation}
  T(\mbox{\boldmath $\theta $}) =\sum_{\mbox{\boldmath $l$}}
  a_{\mbox{\boldmath $l$}} \exp \left( -i\mbox{\boldmath $l \cdot
  \theta $}\right) ,
\end{equation}
where $a_{\mbox{\boldmath $l$}}$ denotes multipole moments. We can
now rewrite the effects of lensing in the multipole expansion:
\begin{eqnarray}
  \widetilde{T}(\mbox{\boldmath $\psi $}) &=&T(\mbox{\boldmath $\psi$ }
  + \delta\mbox{\boldmath $\theta $}) \\
  &=&\sum_{\mbox{\boldmath $l$}}a_{\mbox{\boldmath $l$}}\exp \left[ -i
  \mbox{\boldmath $l \cdot $}(\mbox{\boldmath $\psi $}+\delta
  \mbox{\boldmath $\theta $})\right] , \nonumber
\end{eqnarray}
and
\begin{equation}
  \frac{\partial \widetilde{T}(\mbox{\boldmath $\psi $}^{\prime })}
  {\partial \mbox{\boldmath $\psi $}^{\prime }} =\sum_{\mbox{\boldmath
  $l$}^{\prime }}i \mbox{\boldmath $l$}^{\prime }a_{\mbox{\boldmath
  $l$}^{\prime }}^{\ast } \exp \left[ i\mbox{\boldmath $l$}^{\prime }
  \mbox{\boldmath $\cdot $} (\mbox{\boldmath $\psi $}^{\prime }+
  \delta \mbox{\boldmath{$\theta $}}^{\prime })\right] .
\end{equation}

We can now evaluate the lensing statistic:
\begin{eqnarray}
  H(\xi ) & = & {\cal N} \sum_{\mbox{\boldmath $l$}}
  \sum_{\mbox{\boldmath $l$}^{\prime } }\Bigl \langle
  a_{\mbox{\boldmath $l$}}a_{\mbox{\boldmath $l$}^{\prime }}^{\ast }
  \exp \left( -i \mbox{\boldmath $l \cdot $}\delta\mbox{\boldmath
  $\theta $} +i \mbox{\boldmath $l$}^{\prime }\mbox{\boldmath $\cdot
  $}\delta \mbox{\boldmath $\theta $}^{\prime }\right)
  \label{eqn:Heq} \nonumber \\
  & & \times \left[ i\mbox{\boldmath $l \,\cdot $}\left( \delta
  \mbox{\boldmath $\theta $}_A-\delta \mbox{\boldmath $\theta
  $}_A^{\prime } \right) \right] \exp \left( -i\mbox{\boldmath $l
  \cdot \psi $}+i \mbox{\boldmath $l$}^{\prime }\mbox{\boldmath $\cdot
  \psi $}^{\prime } \right) \Bigr \rangle \\
  & = & {\cal N} \sum_{\mbox{\boldmath $l$}}
  \left\langle a_{\mbox{\boldmath$l$ }}^2
  \right\rangle \exp \left [ -i\mbox{\boldmath $l \,\cdot $}\left(
  \mbox{\boldmath $\psi $}-\mbox{\boldmath $\psi $}^{\prime } \right)
  \right] \nonumber \\
  & & \times \left\langle i\mbox{\boldmath $l \,\cdot $}\left( \delta
  \mbox{\boldmath $ \theta $}_A-\delta \mbox{\boldmath $\theta
  $}_A^{\prime }\right) \exp \left[ -i \mbox{\boldmath $l \,\cdot
  $}\left( \delta \mbox{\boldmath $\theta $}_A- \delta \mbox{\boldmath
  $\theta $}_A^{\prime }\right) \right] \right\rangle \nonumber \\
  & &\times \left\langle \exp \left[ -i\mbox{\boldmath $l \,\cdot$ }
  \left( \delta \mbox{\boldmath $\theta $}_B- \delta \mbox{\boldmath
  $\theta $}_B^{\prime } \right) \right] \right\rangle . \nonumber
\end{eqnarray}
Since the gravitational deflections are the sum of many small
scattering due to superclusters and voids along the line-of-sight, we
can treat $x= \mbox{\boldmath $l$}\cdot \left( \delta\mbox{\boldmath
$\theta $} - \delta \mbox{\boldmath $\theta $}^{\prime }\right) $ as a
Gaussian random variable.
The dispersion of $x$ is
$\left\langle x^2\right\rangle =l^2\sigma ^2$,
where we have kept only the main isotropic term.
The anisotropic term makes a subdominant contribution to the
gravitational lensing effect on two-point auto-correlation function of 
CMB (\cite{sel96,mar97}).
Though the anisotropic term is not too small in this statistics,
the more rigorous analysis will be given in a subsequent paper
(\cite{sug97}).
Then
\begin{eqnarray}
  \left\langle ix\exp (-ix)\right\rangle &=&-\sum_{n=0}^\infty
  \frac{\left\langle (-ix)^{n+1}\right\rangle }{n!}
  \label{eqn:Gauss} \\
  &=&\sum_{n=0}^\infty \frac{(-1)^n}{(2n+1)!}  \left( 2n+1\right) !! 
  \left\langle x^2 \right\rangle^{n+1} \nonumber \\
  &=&l^2\sigma ^2\exp \left( -\frac{l^2\sigma ^2}2\right) . \nonumber
\end{eqnarray}
Combining equations (\ref{eqn:Heq}) and (\ref{eqn:Gauss}),
\begin{eqnarray}
\label{eqn:hxilvec}
  H(\xi ) &=& {\cal N} \sum_{\mbox{\boldmath $l$}} \left\langle
  a_{\mbox{\boldmath $l$}}^2\right\rangle l^2\sigma _A^2(\xi )\exp
  \left[ -\frac{l^2\sigma _A^2(\xi )} 2\right] \exp \left[ -
  i\mbox{\boldmath $l \,\cdot$} \left( \mbox{\boldmath $\psi $}-
  \mbox{\boldmath $\psi $}^{\prime }\right) \right] \nonumber \\
  &&\times \exp \left[ -\frac{l^2\sigma _B^2(\xi )}2\right] .
\end{eqnarray}
The final term represents the smearing of the microwave fluctuations
due to density fluctuations beyond the edge of the redshift
survey. This effect is a small correction term, which we will ignore
for the remainder of the paper.

We average equation (\ref{eqn:hxilvec}) over angle and
rewrite it on the celestial sphere, then
\begin{equation}
  H(\xi )
  = {\cal N} \sum_l C_lW_l\frac{(2l+1)}{4\pi }l^2\sigma _A^2\left
  ( \xi \right) \exp \left[ -\frac{l^2\sigma _A^2(\xi )}2\right]
  P_l(\cos \xi ),
\label{eqn:signal}
\end{equation}
where $\xi $ is the angular separation on the sky, $C_l$ is the
usual multipole moment,
$W_l=\exp(-l^2\sigma_{\rm beam}^2)$ is the window function,
and $\sigma_{\rm beam}$ is the beam of the detector (\cite{kno95}).

The cosmic variance in the cross-correlation statistic can also be
estimated by taking the leading order term in the expansion:
\begin{eqnarray}
  \left\langle H^2\right\rangle &=& {\cal N}^2 \left\langle \left[
  \widetilde{T}( \mbox{\boldmath $\psi $}) \left( \left[ \delta
  \mbox{\boldmath $\theta $} -\delta \mbox{\boldmath $\theta
  $}^{\prime } \right] \mbox{\boldmath $\cdot \nabla $}\widetilde{T}(
  \mbox{\boldmath $\psi $}^{\prime })\right) \right] ^2\right\rangle
  \\
  &\simeq &{\cal N}^2 \sigma _A^2\left( \xi \right) \left\langle
  T^2\right\rangle \left\langle \left( \nabla T\right)
  ^2\right\rangle. \nonumber
\end{eqnarray}
Note that the signal-to-cosmic variance scales as $^{}\left\langle
H\right\rangle /$ $\left\langle H^2\right\rangle ^{1/2},$ which is
proportional to $\sigma _A.$ As we claimed earlier, this is a
first-order effect.

We can estimate the signal-to-cosmic variance ratio by computing the
predicted signal per multipole:
\begin{eqnarray}
  H_{\mbox{\boldmath $l$}} &=&\int
  d\mbox{\boldmath$\psi$}d\mbox{\boldmath $\psi $}^{\prime }
  \widetilde{T_{\mbox{\boldmath $l$}}}(\mbox{\boldmath $\psi$}) \left
  ( \delta \mbox{\boldmath $\theta $} (\mbox{\boldmath $\psi $})-
  \delta \mbox{\boldmath $\theta $}^{\prime}(\mbox{\boldmath
  $\psi$}^{\prime }) \right) \mbox{\boldmath $\cdot \nabla $}
  \widetilde{T_{\mbox{\boldmath $l$}}}^{\ast } (\mbox{\boldmath $\psi$
  }^{\prime }) \nonumber \\
  &=&\int d\mbox{\boldmath $\psi $}d\mbox{\boldmath $\psi $}^{\prime }
  a_{\mbox{\boldmath $l$}}\exp \left[ -i\mbox{\boldmath $l \,\cdot $}
  \left( \mbox{\boldmath $\psi $}+\delta\mbox{\boldmath $\theta$}
  \right) \right] \left[ i \mbox{\boldmath $l \,\cdot $}\left( \delta
  \mbox{\boldmath $\theta $}( \mbox{\boldmath $\psi $})-\delta
  \mbox{\boldmath $\theta$}^{\prime}(\mbox{\boldmath $\psi $}^{\prime })
  \right) \right] \nonumber \\
  &&\times a_{\mbox{\boldmath $l$}}^{\ast } \exp \left[
  i\mbox{\boldmath $l \,\cdot$} \left( \mbox{\boldmath $\psi
  $}^{\prime }+\delta\mbox{\boldmath $\theta $} ^{\prime }\right)
  \right] \\
  &\simeq &C_lW_l\,l^2\sigma _l^2\exp (-l^2\sigma _l^2/2), \nonumber
\end{eqnarray}
where $\sigma _l=\sigma _A(\pi /l).$ If we include detector noise,
then
\begin{eqnarray}
  \left\langle T_l^2\right\rangle &=&C_lW_l+w^{-1}, \\
  \left\langle \left( \nabla T\right) _l^2\right\rangle \nonumber
  &=&l^2\left( C_lW_l+w^{-1} \right), \nonumber
\end{eqnarray}
where $w^{-1}$
is the measure of the detector noise (\cite{kno95}).  \ MAP's highest
frequency channel has a full width at half maximum of $0.21 \,{\rm
degree}$ and a system noise of $ w^{-1}=\left( 10.5\mu {\rm K}\right)
^2\,{\rm degree}^2 =(0.18\mu {\rm K})^2\,{\rm steradian}$. By
combining the three highest frequency channel, the MAP\ system noise
drop to $(0.11\mu {\rm K})^2\,{\rm steradian}$.  This implies that the
noise plus cosmic variance in $ H_l$ is
\begin{equation}
  \left\langle \left( H_l\right) ^2\right\rangle
  =\frac{l^2\sigma_l
  ^2}{2l+1 }\left( C_lW_l +  w^{-1} \right)^2,
\end{equation}
where the factor of $2l+1$ comes from averaging over all multipoles
with the same $l.$ Note that there is no factor of 2 in the numerator
as the variance is proportional to the product of two uncorrelated
fields, $ \left\langle T^2\left( \nabla T\right) ^2\right\rangle $,
rather than the more familiar $\left\langle T^4\right\rangle -$
$\left\langle T^2\right\rangle ^2=2$ $\left\langle T^2\right\rangle
^2.$ Summing over all multipoles yields the expected signal in the
microwave maps:
\begin{eqnarray}
  \label{eqn:snr}
  \left( \frac SN\right) ^2 &=&\sum_l\frac{\left\langle H_l
  \right\rangle ^2}{\left\langle \left( H_l\right) ^2\right\rangle }
  \\
  &=&\sum_l(2l+1)\frac{l^2\sigma _l^2\exp \left( -l^2\sigma _l^2
  \right) } {\left[ 1+\left( wC_lW_l \right)^{-1} \right] ^2}.
  \nonumber
\end{eqnarray}

Note that an additional cross-correlation function that can be
computed from the microwave background fluctuations and the lensing
maps:
\begin{equation}
  G\left( \xi \right) =\left\langle \widetilde{T}(\mbox{\boldmath
  $\psi $} )\left( \left[ \delta\mbox{\boldmath $\theta $} _
  A(\mbox{\boldmath $\psi $})- \delta\mbox{\boldmath $\theta $}^{\prime}_
  A(\mbox{\boldmath $\psi $}^{\prime })\right] \mbox{\boldmath $\cdot
  \,$}\mbox{\boldmath $\widehat{n}$} \widetilde{T}(\mbox{\boldmath
  $\psi $}^{\prime }) \right) \right\rangle _\xi ,
\end{equation}
where $\mbox{\boldmath $\widehat{n}$}= \frac{\left( \mbox{\boldmath
$\psi $}-\mbox{\boldmath $\psi $} ^{\prime }\right) }{\left|
\mbox{\boldmath $\psi $}-\mbox{\boldmath $\psi $} ^{\prime }
\right|}.$
The statistics has the simplest form among possible inclusions of
gravitational lensing angular excursion.
The cross-correlation function is reduced to
\begin{equation}
  G\left( \xi \right)
  = {\cal N}_G \sigma_A^2 \frac{\partial C(\xi)}{\partial \xi}.
\end{equation}
However we confirmed that the expected signal-to-noise ratio,
$S/N$, in $G$ is lower than in $H$.
This is because, unlike in $H$, the cosmic variance in $G$ contains a
term proportional to
$\langle T^2(\mbox{\boldmath $\psi$})
T^2(\mbox{\boldmath $\psi$}^{\prime}) \rangle$.

%
%

\section{Results and Discussion}

We have computed the expected cross-correlation between the
temperature fluctuations and the lensing bending angle (equation
(\ref{eqn:signal})) for the standard CDM\ model and the ``best fit''
vacuum dominated model (\cite{ost95}).  In estimating the lensing
bending angle, we have assumed that we have a redshift survey that
extends to a characteristic redshift $z.$ Figure 2 shows the results
for large scale surveys of varying depths. In this figure, we have
assumed the standard parameters for the MAP\ and SDSS\ projects. The
characteristic depth of the SDSS\ redshift survey is $z=0.2.$ SDSS\
has photometric redshifts for 10$^8$ galaxies that should extend the
survey to $z=1.$ These photometric redshifts are accurate to $\pm
0.02$ in redshift; certainly accurate enough to compute the projected
surface densities needed to predict lensing. Figure 3 shows the
cumulative signal-to-noise, where we have summed over all of the
multipoles.  We have divided $(S/N)$ in equation (\ref{eqn:snr}) by a
factor of $2$ taking into account the limited sky coverage, $\Omega _
{\rm SDSS}$.  The predicted signal-to-noise is quite large (15 and 35)
for the vacuum-dominated and standard CDM\ models.

%
\begin{figure}[t]
  \plotone{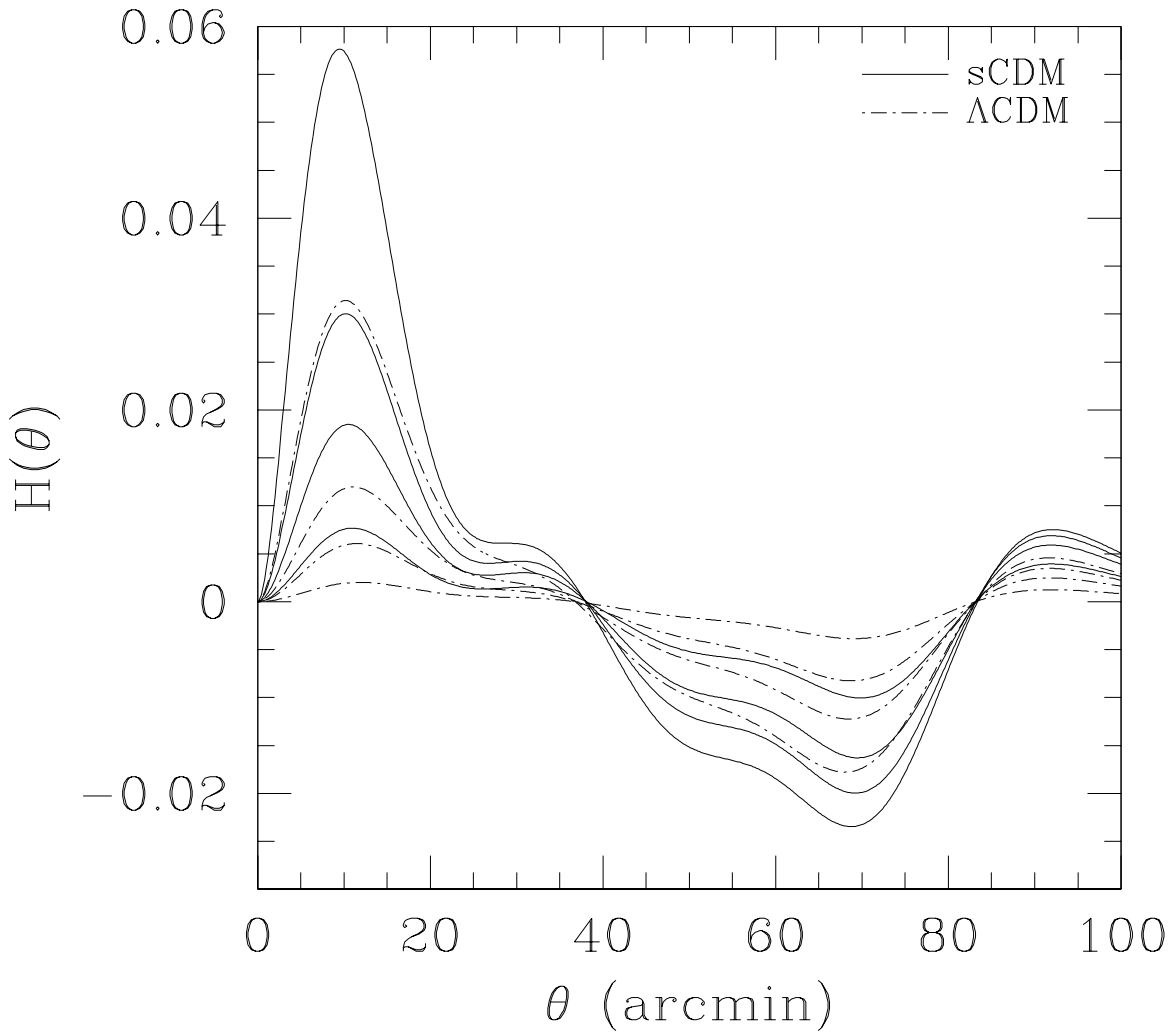}
  \caption{The cross-correlation, $H(\theta)$, for
     $\theta_{\rm fwhm}=0.21\, {\rm degree}$.  Solid and broken lines
     show the SCDM model and the $\Lambda$CDM model as in Fig.~1.  For
     each model, the curves show the case in which $z = z_{\rm dec}$,
     $1$, $0.5$, and $0.2$.}
  \label{fig02}
\end{figure}
%
%
\begin{figure}[t]
  \plotone{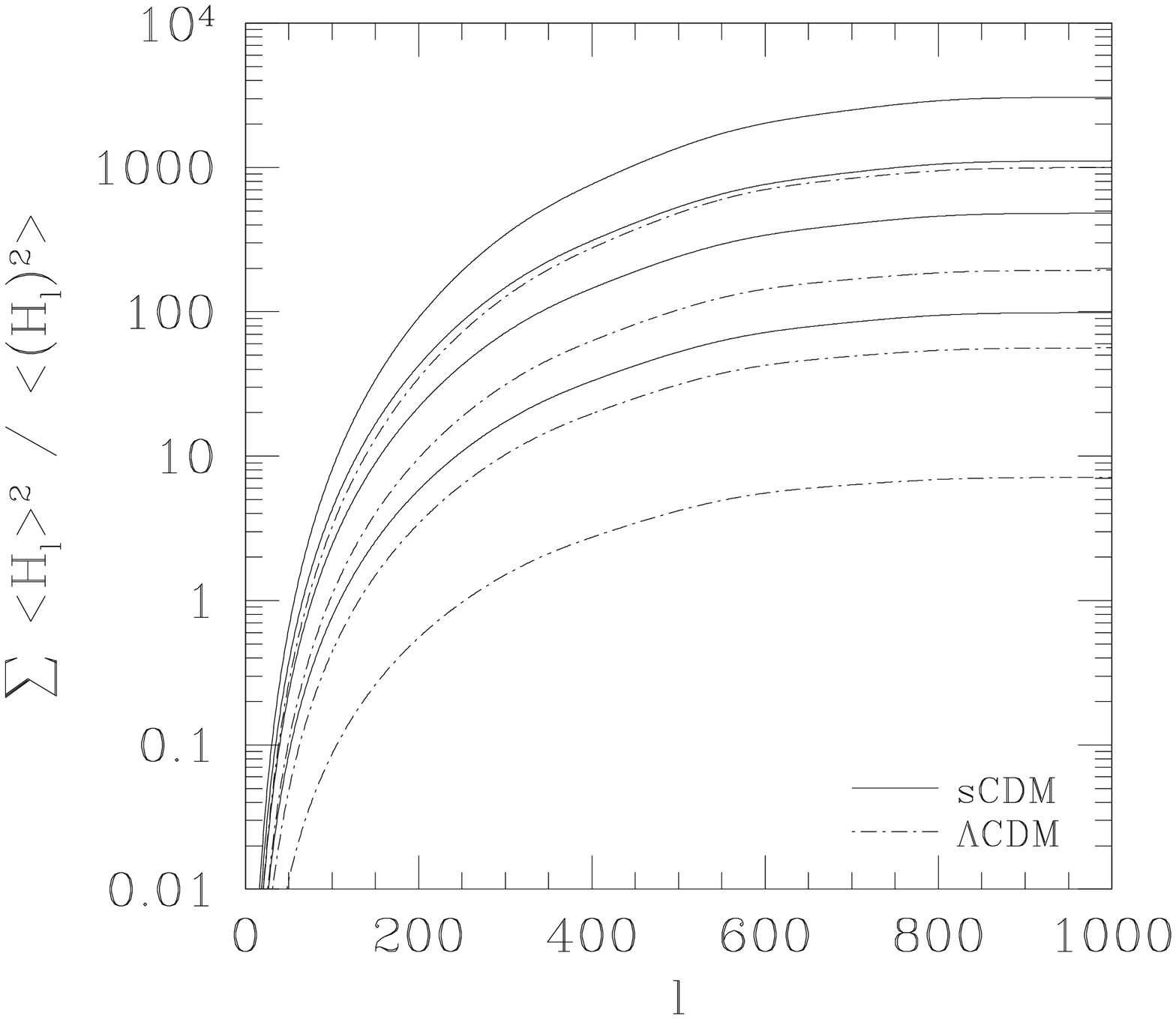}
  \caption{The cumulative signal-to-noise ratio.  Solid
     and broken lines show the SCDM model and the $\Lambda$CDM model
     as in Fig.~1.  For each model, the curves show the case in which
     $z = z_{\rm dec}$, $1$, $0.5$, and $0.2$.}
  \label{fig03}
\end{figure}

This cross-correlation, if detected, directly probes the gravitational
potential fluctuations at low redshift. In principle, it should yield
an accurate determination of the biasing factor. Armed with this
measurement, we should be able to directly compare the gravitational
potential fluctuations at decoupling with the gravitational potential
fluctuations in the local universe.

\acknowledgments
 We thank Seljak and Zaldarriaga for providing their code to generate
the intrinsic CMB power spectrum.  DNS acknowledges the MAP/MIDEX
project for support.  MS and TS acknowledge support from Research
Fellowships of the Japan Society for the Promotion of Science.

\clearpage
%
%

%
\end{document}